\documentclass[usenatbib]{basi}
\usepackage[T1]{fontenc}
\usepackage[british]{babel}
\usepackage[varg]{txfonts}
\usepackage{rotating}
\usepackage{dcolumn}
\def\degr{\hbox{$^\circ$}}
\begin{document}
\title[A new astronomical polarimeter]{A new three-band, two beam astronomical 
photo-polarimeter}

\author[G. Srinivasulu et~al.]
{G. Srinivasulu$^1$\thanks{email: \texttt{gs@iiap.res.in}},
 A. V. Raveendran$^2$,
 S. Muneer$^1$\thanks{email: \texttt{muneers@iiap.res.in}},
 M. V. Mekkaden$^3$, 
\newauthor
 N. Jayavel$^4$,
 M. R. Somashekar$^1$,
 K. Sagayanathan$^1$,
 S. Ramamoorthy$^1$,
\newauthor
 M. J. Rosario$^5$ and
 K. Jayakumar$^6$\thanks{email: \texttt{jayakumar.kanniah@gmail.com}}\\
 $^1$Indian Institute of Astrophysics, Bangalore~560034, India\\
 $^2$399, "Shravanam", 2$^{nd}$ Block, 9$^{th}$ Phase, J P Nagar, 
Bangalore~560108, India\\
 $^3$No 82, 17E Main, 6$^{th}$ Block, Koramangala, Bangalore~560095, India\\
 $^4$No 22, Bandappa lane, New Byappanahalli, Bangalore~560038, India\\
 $^5$210, 4$^{th}$ Main, Lakshmi Nagar Extn, Porur, Chennai~600116, India\\
 $^6$24, Postal Nagar, Ampuram, Vellore-632009, India}

\pubyear{2014}
\volume{42}
\pagerange{\pageref{firstpage}--\pageref{lastpage}}

\date{Received 2014 July 23; revised    ;  accepted}

\maketitle

\label{firstpage}

\begin{abstract}
We designed and built a new astronomical photo-polarimeter that
can measure linear polarization simultaneously in three spectral bands.
It has a Calcite beam-displacement prism as the analyzer. The ordinary
and extra-ordinary emerging beams in each spectral bands are 
quasi-simultaneously detected by the same photomultiplier by using a high speed
rotating chopper.
 A rotating superachromatic Pancharatnam halfwave
plate is used to modulate the light incident on the analyzer. The spectral
bands are isolated using appropriate dichroic and glass filters.

 We show that the reduction of 50\% in the efficiency of
 the polarimeter because of the fact that the intensities of the two beams
are measured alternately is partly compensated by the reduced time to be
spent on the observation of the sky background. The use of a beam-displacement
 prism as the analyzer completely removes the polarization of 
background skylight, which
is a major source of error during moonlit nights, especially, in
the case of faint stars.

The field trials that were carried out by observing several
polarized and unpolarized stars show the performance of the polarimeter
to be satisfactory.
\end{abstract}


\section{Introduction}\label{s:intro}
There was a need for an efficient photo-polarimeter for observations with the
1-m Carl Zeiss Telescope at Vainu Bappu Observatory, Kavalur. The single
spectral band, star--sky
chopping polarimeter, which was built by \citet{jain} almost a quarter century
ago, was the only available instrument for polarimetric studies
\citep{asho, part, mano, rave99, rave02}.
The star-sky chopping procedure makes the instrument very inefficient,
thereby underutilizing the available telescope time.

A project to build a new
photo-polarimeter for observations of point sources 
was initiated in the Indian Institute of Astrophysics
quite sometime back. Unfortunately, due to unforeseen reasons there
were delays at various stages of the execution of the project.
There were several ongoing programmes
in the Institute which required accurate multiband polarimetry:
studies of Herbig Ae/Be stars \citep{asho},
Luminous Blue variables \citep{part},
Young Stellar Objects \citep{mano}, R CrB stars \citep{kame},
RV Tauri stars \citep{rave99}, Mira
variables that show long-term modulations in their brightness at light maximum
\citep{rave02} and T~Tauri stars \citep{mekk}.

The new polarimeter, which was designed and built in the Institute,
 was mounted onto the 1-m telescope and observations of several polarized and unpolarized
stars were made during April--May~2014, to determine its suitability for efficient astronomical
observations. Due to prevalent poor sky conditions, the instrument could be
used effectively only on a few nights during this period. The instrument was
found to have a very high mechanical stability, but 
a comparatively low polarization efficiency of 94.72\%. Since we suspected
the scattered light inside the instrument to be the reason for such a low value
for the efficiency, we blackened the few internal mechanical parts which were
left out earlier, and again made observations during February--April~2015. 

In this paper we discuss the details that have gone in the selection of the
analyzer and give brief descriptions of the component layout, the control 
electronics, and the data acquisition procedure and analysis.
Details of the polarimeter and the data reduction procedure
can be found in \citet{rave15}.
We also discuss the results of the observation made during 
February--April~2015, which were carried out to 
evaluate the performance of the instrument.

\section {Choice of the analyzer}
In order to reduce the adverse effects of the atmospheric scintillation, seeing
and transparency variation, which usually limit the accuracy of any
polarimetric measurement, it is imperative to measure 
simultaneously both the emerging beams,
ordinary and extra-ordinary, from the analyzer. The two basically different
types of analyzers that are usually employed in astronomical polarimeters are:
(1) beam-displacement prisms made of Calcite \citep{piir, maga88, scal, schw}
and (2) beam-splitting prisms of Wollaston or Foster design
 \citep{maga84, desh, kiku, houg}

\begin{figure}
\centerline {\includegraphics[width=8.8cm]{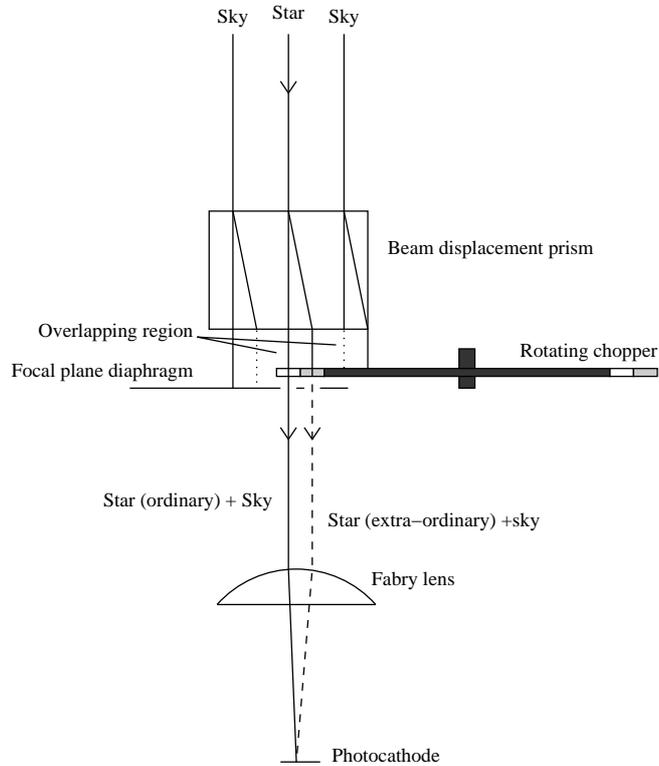}}
\caption {Working principle of the polarimeter.\label{f:wpp}}
\end{figure}
We adopted a design for the polarimeter incorporating a Calcite
beam-displacement prism so that most of the spectroscopic nights
that are available at the site can be utilized for polarimetric observations.
The working principle of the polarimeter is illustrated in Fig.~\ref{f:wpp}.
An astronomical polarimeter based on this principle was first built
by \citet{piir}.
A beam-displacement prism divides the incident
light, producing two spatially separated emerging beams
with mutually perpendicular planes of polarization.
The background sky, which acts as an extended object, illuminates
the entire top surface of the
beam-displacement prism, and hence, produces two broad emergent
beams whose centres are spatially separated.
There will be considerable overlap between these two beams
about the geometrical axis of the prism, and wherever the
beams overlap in the focal plane of the telescope that portion remains
unpolarized by the prism and gives the background
sky brightness directly.
The situation is different with respect to the starlight;
the star being a point source, there will be no overlap between the
two emergent beams from the prism. The net result of
these two effects is that we observe two images of the star
with mutually
perpendicular planes of polarization at the focal plane, which are
superposed on
the unpolarized background sky. Two identical apertures are used to
isolate these images, and a rotating chopper is used to alternately
block one of the images allowing the other to be detected by the
same photomultiplier tube.

The main advantages of such an arrangement are:
(i) The contribution of background sky polarization is completely eliminated
from the data, thereby, facilitating the observations of fainter stars during
moonlit nights without compromising on the accuracy that is achievable
during dark nights. This is possible because the background sky
is not modulated, it just appears
as a constant term that can be removed from the data accurately.
(ii) Since the same photomultiplier tube is used to
detect both the beams quasi-simultaneously, the effect of any time-dependent
variations in its sensitivity
as a result of undesired
variations in the operating conditions of the
associated electronics, like, variations in HT supply,
is negligible.
(iii) The quasi-simultaneous detection of both the beams using a fast
rotating chopper essentially eliminates the adverse effects of the variations
in sky-transparency and significantly reduces the errors due to
atmospheric scintillation for bright stars.
Scintillation noise is independent of the brightness of the star
and dominates the photon noise for bright objects; with a 1-m
telescope the low frequency ($<$ 50~Hz) scintillation noise is
expected to be larger than the photon noise for stars brighter
than B~$=$~7.0~mag at an airmass of 1.0 \citep{youn}.
The averaging of the data by the process of long integration will reduce
the scintillation noise
only to a certain extent because of its log-normal
distribution. The frequency spectrum of scintillation
noise is flat up to about 50~Hz; the noise amplitude decreases rapidly
above this frequency and it becomes negligibly small above 250~Hz.
With the fast chopping of the two
emergent beams alternately and the automatic removal of the background
sky polarization, it is possible to make polarization
measurements that are essentially photon-noise limited.

The two factors that determine the overall efficiency of a polarimeter
attached to a telescope are \citep{serk1}:
(i) the faintest magnitude that could be reached with a specified
accuracy in a given time, and (ii) the maximum amount of information
on wavelength dependence that could be obtained during the same time.
It is evident that
the former depends on the efficiency in the
utilization of photons collected
by the telescope and the latter on the number of spectral bands
that are simultaneously available for observation.
In the beam-displacement prism-based polarimeters, where the image
separation is small, the intensities
of the two beams produced by the analyzer are measured alternately
using the same detector, and hence, only 50 per cent of the light
collected by the telescope is effectively utilized.
In polarimeters using Wollaston prism
as the analyzer, the two beams, which are well-separated
without any overlap, can be detected simultaneously by two
independent photomultiplier tubes, fully utilizing
the light collected \citep{maga84, desh}.
For simultaneous multi-spectral band
observations that make use of the incident light fully,
the beams emerging from the analyzer will have to be well-separated so as
to accommodate a large number of photomultiplier tubes,
making the resulting instrument both heavy and large in size \citep{serk},
and hence, not suitable for the 1-m telescope.
Usually, in Wollaston and Foster
prism-based polarimeters, which have the provisions for
simultaneous observations in multi-spectral bands,
different spectral bands are distributed
among the two emergent beams, and hence,
at a time only 50 per cent of the light
collected in each spectral region is made use of \citep{kiku, houg}.

The background sky polarization has to be determined accurately and
removed from the object $plus$ sky data
when there is no overlap between the emergent beams, and a significant
fraction of the time spent on object integration will have to be
spent for such observations if the objects are faint.
In beam-displacement prism-based polarimeters, time has to be spent
only to determine the brightness of the background sky, and
for the maximum signal-to-noise ratio, it is only a negligible
fraction of the object integration
time even for relatively faint objects; the available time for observation
can be almost entirely utilized
to observe the object, as indicated by the
analysis presented in section \ref{ss:optsky} on optimum
sky background observation.
Therefore, the non-utilization of 50 per cent
light in beam-displacement prism-based polarimeters does not
effectively reduce
their efficiency much, 
especially, while observing faint objects where the
photons lost actually matter. The provisions for multi-spectral
band observations can be easily incorporated in the design, making
beam-displacement prism-based polarimeters to have an overall
efficiency significantly higher than that of the usual
Wollaston or Foster prism based polarimeters.
Another advantage of
these polarimeters is that they can be easily converted into
conventional photometers,
just by pulling the beam-displacement prism out of the light path.

\begin{figure}
\centerline{\includegraphics[width=9.5cm]{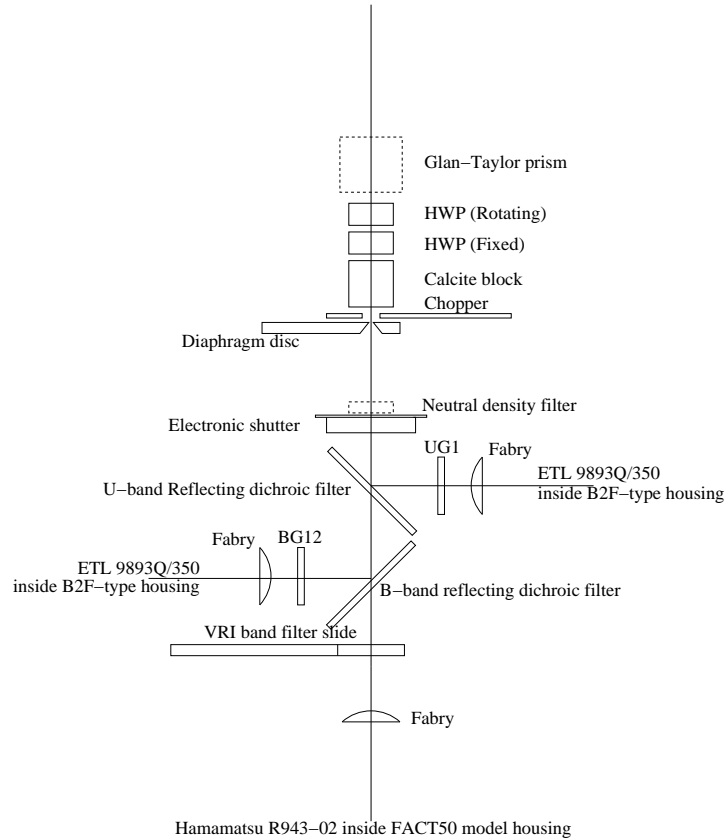}}
\caption {Schematic layout of the main components of the polarimeter.
In the instrument the $U$ and $B$ bands are mutually perpendicular,
making it more compact.
\label{f:slo}}
\end{figure}

\begin{figure}
\centerline {\includegraphics[width=9.2cm]{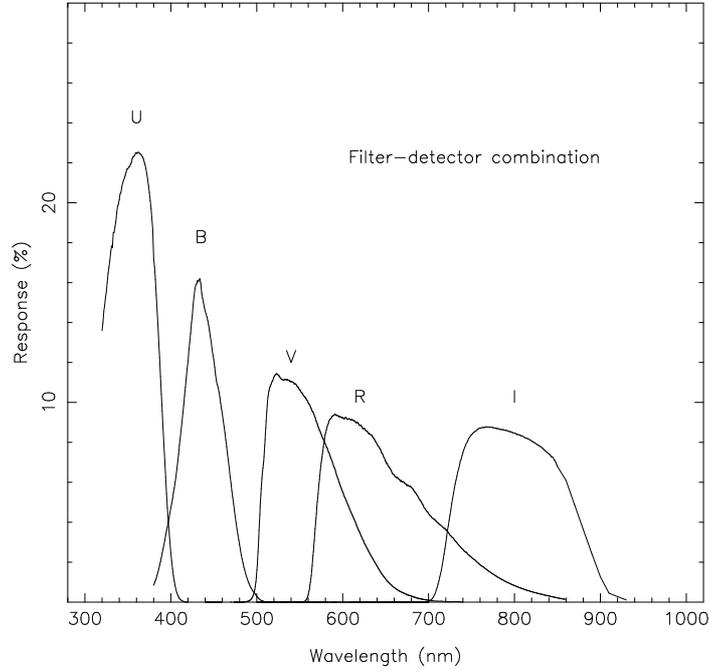}}
\caption {Combined response of the filter-detector combination.
\label{f:crfd}}
\end{figure}
\section{Layout of the components}
The layout of the polarimeter indicating the positions of the main
components is shown schematically in Fig.~\ref{f:slo}. 
A super-achromatic
halfwave plate of Pancharatnam design is used to rotate the
plane of polarization of the incoming light falling on the analyzer,
and thereby, to modulate the intensities of the two emergent beams
from the analyzer \citep{frec}.
The position angle of the effective optical axis of a Pancharatnam retarder
is wavelength-dependent. 
This would cause difficulties in the accurate determination of the
position angle of polarization when broad spectral bands, as in the
present case, are used for observations because corrections that are to be
applied to the observed
position angles would depend on the energy distribution
of the observed objects. 
Such an inconvenience is usually avoided
by introducing another stationary identical Pancharatnam plate
immediately after the rotating plate in the light path. 
The introduction of such a half-wave plate
ensures that the signal modulation is only a function of the relative
positions of the effective optical axes of the two half-wave plates,
which is wavelength-independent.
The rotating halfwave 
plate, which acts as the modulator of the incoming light, is the first
element in the optical path of the polarimeter, and hence, avoids the
detection of any spurious
polarization produced by any optical components used in the polarimeter.  
The two identical
half-wave plates of 19~mm clear aperture
used in the polarimeter are acquired
from Bernhard Halle Nachfl. GmbH, Berlin, Germany.

The analyzer
is a cross-mounted, double calcite block with a 20~mm clear aperture and a
total thickness of 28~mm, which produces identical ordinary
and extraordinary images at the focal plane of the telescope. 
This component is also acquired from Bernhard Halle Nachfl. GmbH.
Since the
separation between the emergent beams is a function of wavelength, the
images will be slightly dispersed. The image strips
have a linear size of 0.318~mm in the spectral interval 320--990~nm with the
centre-to-centre distance of the two images being 2.292~mm. The images
can be isolated for observation using one of the two sets of twin
identical diaphragms of 20 and 25~arc sec in diameter mounted on 
a manually rotatable disc. The
calcite block is placed
in the light path immediately after the stationary half-wave plate. 
The rotating chopper isolates alternately the ordinary or extra-ordinary
image for detection by the same photomultiplier. The chopper has four slots,
two for each of the images, and therefore
the
effective chopping frequency is double the rotational frequency of
the chopper.

\newcolumntype{d}[1]{D{.}{.}{#1}}
\begin{table}
\caption{Filter-detector combinations.}\label{t:fpmt}
\medskip
\begin{center}
\begin{tabular}{cccc}\hline
Spectral &  & Mean wavelength & \\
Band & Filter combination & $\lambda_0$\,(nm) & Photomultiplier tube\\
\hline
U  & UG1 (1~mm) & 357 & ETL 9893Q/350B\\
B  & BG12 (1~mm) & 437 & ETL 9893Q/350B\\
V  & BG18 (1~mm) + GG495 (2~mm) & 561 & Hamamatsu R943-02\\
R  & OG570 (2~mm) + KG3 (2~mm) & 652 & $_{''}$\\
I  & RG9 (3~mm) & 801 & $_{''}$\\
\hline
\end{tabular}\\[5pt]
\end{center}
\end{table}
The isolation of the spectral regions into the three bands
for simultaneous observation is achieved
by using two dichroic filters, one to reflect the ultra-violet
part of the
incoming light and the other to reflect
the blue part of the spectrum from the light transmitted by the first.
The dichroic filters used are obtained from Custom Scientific, Inc.,
Arizona, USA.
The reflective coatings required for the dichroism
are done on 2~mm thick glass substrates.
The two reflected beams are detected by two separate uncooled
photomultiplier tubes.
The bi-alkali photocathodes of these tubes along with
the Schott glass filters
inserted in front of them produce
spectral bands that approximate the $U$ and $B$\, bands of Johnson.
The light transmitted by both the dichroic filters, which
fall in the $VRI$ spectral bands, is detected by a
cooled photomultiplier tube with a GaAs photocathode. The observations are
made sequentially in $V$, $R$ and $I$ bands using suitable filter combinations
mounted on a sliding filter-holder. The
Schott glass filters that are used have clear apertures of
25~mm diameter.
The combined responses of the
dichroic mirrors, glass filters and detectors are
plotted in Fig.~\ref{f:crfd}. 
The $V$ passband approximates that of Johnson's, while the
$R$ and $I$ passbands approximate those of Cousins \citep{bess79, bess93}.
The selection of the required glass filters are based
on the filter-detector combinations used by \citet{piir} and \citet{houg};
these authors also have employed dichroic
filters to isolate spectral regions in their multiband polarimeters.
The details of the filter-detector combinations used are given in
Table~\ref{t:fpmt}. The mean wavelength, calculated using
$$ \lambda_0 = \frac{\int \lambda\,S(\lambda )\,\delta\lambda}
{\int S(\lambda )\,\delta\lambda},$$ 
is also given in the table against the respective
spectral band; S($\lambda$) is the wavelength-dependent
responses plotted in Fig.~\ref{f:crfd}.

The pulse amplifier-discriminators of Model ETL AD6 and Hamamatsu
Model C9744 are used to interface the ETL~9893Q/350 and Hamamatsu R943-02
photomultiplier tubes, respectively, to the pulse counters.
The uncooled ETL 9893Q/350 photomultiplier tubes used
in the $U$ and $B$\, bands give dark counts which are
 less than 10 counts~s$^{-1}$
at an ambient temperature of about 25\degr C when operated at an anode voltage
of 2300~V. These are mounted in uncooled ambient ETL B2F-type
housings, while the Hamamatsu tube used in the $VRI$ channel is mounted in a 
FACT~50 model housing, which is of forced air-cooled type and cools 
about 50\degr C
below an ambient temperature of 20\degr C. The dark counts given by the cooled
tube when operated at a cathode voltage of --1900~V are also less than 
10 counts~s$^{-1}$.

A Glan-Taylor prism of 24.5 mm clear aperture, also acquired from
Bernhard Halle Nachfl. GmbH, can be inserted in the telescope light
path to produce
a fully polarized beam that is necessary to 
measure and periodically
check out the constancy of the polarization efficiency
of the instrument.
Facilities for wide angle-viewing and diaphragm-viewing have also been
provided in the
instrument for the easy acquisition of the object for observation.

\section{ Control electronics}
All the polarimeter functions are performed by two PIC family
PIC16F877A1 microcontrollers. 
One microcontroller controls  both the step motor
coupled to the half wave plate and the servo motor coupled to the
chopper wheel. The second microcontroller interfaces with the Linux machine,
which is used to operate the polarimeter,
through the standard serial communication with the RS232C protocol. 
The two microcontrollers in the instrument 
communicate between them through the built-in SPI Bus.

We have used a command-based communication protocol, 
wherein the computer acts as the master and the microcontroller as a slave
unit.  Always the commands are generated from the computer and sent to the
microcontrollers, which
interpret and execute them, and then respond indicating the
status of the execution to the computer. 
We followed this scheme to avoid any communication clash between the computer
and the microcontrollers, and to identify the errors, if any, quickly.

\section{Data acquisition and reduction procedure}
\subsection {Observations}
The photomultiplier pulses corresponding to the
intensities of the two emergent beams from the beam-displacement prism
are counted separately 
over a full rotation of the halfwave plate at the specified
equal angular intervals, starting from a reference position.
The actual counting time interval for each beam over a rotational
cycle of the chopper depends on its
frequency of rotation since the latching pulses for the electronic pulse
counters are derived from the positional sensors of the chopper.
When the integrated counting
time over several rotational cycles of the chopper
equals to what is specified, the counting is stopped, and the resulting
counts are stored. The halfwave plate is then moved to its next position, and
the process is repeated at all the required positions.
 The entire
procedure is counted as one cycle. The cycle can be repeated till the required
accuracy in the measurement is achieved. After each cycle, the counts are
added to the previously stored counts at the respective position of the
halfwave plate.  Before a new cycle begins
 the halfwave plate is always brought to its reference position.
In order to give equal weightage to the observations in the data reduction,
the number of counts accumulated at all positions of the half-wave
plate should be of the same order. This requires that under poor sky
conditions the observations should be
repeated over several cycles of rotation of the half-wave plate, with a
smaller integration time at each position of the halfwave plate.
Usually, the observations in $U$ and $B$
bands will last longer than those in $V$, $R$ and $I$ bands.
The integrations in $U$ and $B$ can be continued till integrations in
the other bands are completed successively.

The procedure is the same for the object and the sky background
that has to removed from the object $plus$ sky background
counts before the data reduction.
As shown in section \ref{ss:optsky} on optimum background sky observation,
 the time to be spent on sky integration is usually a small fraction of the
time spent on the object integration. The optimum number of sky cycles for
an observed number of object cycles and the optimum number of object cycles
for an observed number of sky cycles, which are computed from the relative
brightnesses of the object and sky, are displayed on the monitor.

After each cycle, the linear polarization, the position angle
and the gain-ratio, and
their probable errors are displayed on the monitor. If the sky background is
observed first, the displayed values will represent the actual values, otherwise
they will only be approximations. Once the integration of the object is
terminated, the final values of the Stokes parameters $Q$ and $U$,
the polarization (P\%), the position angle ($\theta$\degr)
and the mean Julian day of observation
are stored in appropriate files.

\subsection{Data reduction}
The pair of counts registered at each position $i$
of the halfwave plate depends
both on
the Stokes parameters $I$, $Q$ and $U$ of the incident light and the angle
$\psi^i$ between the optical axes of the two halfwave plates at that
position as given below:
\begin{eqnarray}
&N_1^i = \frac{1}{2}\,G_1\,(I + Q\,\cos 4\psi^i + U\,\sin 4\psi^i)\,T_{atm}
\quad \mbox{and} \nonumber \\
&N_2^i = \frac{1}{2}\,G_2\,(I - Q\,\cos 4\psi^i - U\,\sin 4\psi^i)\,T_{atm}\, ,\nonumber
\end{eqnarray}
where the subscripts 1 and 2 refer to the beams polarized in the
principal plane of the first plate of the calcite block and the plane
perpendicular to it, respectively. The parameters $G_1$ and $G_2$
are the net efficiencies for the two light beams, which include the
transmittances and reflectivities of the various optical elements in the
light path and the quantum efficiencies of the cathode of the photomultiplier
tube, and $T_{atm}$ is the atmospheric transmittance integrated over
the period of observation.
Taking the ratio of the counts accumulated in the two channels,
\begin{eqnarray}
 Z^i = \frac{N_1^i}{N_2^i} = \frac{1 + q\,\cos 4\psi^i + u\,\sin 4\psi^i}
{\alpha\,(1 - q\,\cos 4\psi^i - u\,\sin 4\psi^i)}\, , \nonumber
\end{eqnarray}
where $\alpha$ ($=G_2/G_1$) is the ratio of the efficiencies
of the two channels (gain-ratio),
 and $q$ and $u$ are the normalized Stokes parameters,
($Q/I$) and ($U/I$). 

There will be $M$ such ratios corresponding to the $M$ positions of the
halfwave plate over its full rotation
at which the photons in the two beams are
counted. The Stokes
parameters and the gain-ratio are solved using the least-square
method from these $M$ ratios.
The normal equations are solved using the Cracovian matrix elimination method
\citep{kopa} because it also gives the weights required to calculate the 
probable
errors in the parameters that are solved for from the 
resulting standard deviation of the
least square fit. 

The percentage linear polarization P, position angle $\theta$  and their
probable errors, $\epsilon_P$ and $\epsilon_\theta$, are calculated from
\begin{eqnarray}
P\,(\%) = \sqrt{q^2 + u^2} \times 100\, , \nonumber\\
2\,\theta = \tan^{-1}(u/q)\, , \nonumber\\
\epsilon_P\,(\%) = 
\frac{\sqrt{(q\,\epsilon_q)^2 + (u\,\epsilon_u)^2}}{p} \times 100 
\quad \mbox{and}  \nonumber\\
\epsilon_\theta \,\mbox{(degree)} =
 \frac{28.65\,\sqrt{(q\,\epsilon_u)^2 + (u\,\epsilon_q)^2}}{p^2}\, ,\nonumber
\end{eqnarray}
where $\epsilon_q$ and $\epsilon_u$ are the probable errors in $q$ and $u$,
determined from the least square solution.

\subsection{Optimum background sky observation}\label{ss:optsky}
As already mentioned,
the time to be spent on sky integration is usually a small fraction of the
time spent on the object integration for the minimum error in the polarization
determined, which arise from the statistical fluctuations in the photons
counted. If $n_1^i$ and $n_2^i$ are the observed
star $plus$ background sky counts at the $i^{th}$ position of the halfwave
plate,
and $s_1$ and $s_2$ are the scaled background sky counts due to the two beams,
\begin{eqnarray}
N_1^i =  n_1^i - s_1 \, \, \mbox{and} \, \, N_2^i = n_2^i - s_2. \nonumber
\end{eqnarray}
If we assume the gain-ratio to be unity, which is expected to be close to unity
in normal cases, the conditions for least square give
\begin{eqnarray}
q  =&& \frac{\sum(Z_i^2 -1)\, \cos 4\psi_i\, \sum(1 + Z_i)^2\, \sin^24\psi_i -
 \sum(Z_i^2 -1)\, \sin 4\psi_i\, \sum(1 + Z_i)^2\, \sin 4\psi_i\, \cos 4\psi_i}
  {\sum(1 + Z_i)^2\, \cos^24\psi_i\, \sum(1 + Z_i)^2\, \sin^24\psi_i - 
   (\sum(1 + Z_i)^2\, \sin 4\psi_i\, \cos 4\psi_i)^2}\, . \nonumber 
\end{eqnarray}
Further, if we assume that the observed polarization is small so that the
 squares and
higher order terms in $q$ and $u$ can be neglected, and that the measurements
are done over a full rotation of the halfwave plate at equal angular
intervals, 
the expression for the differential in $q$ can be written as
\begin{eqnarray}
 M\, n_*\, \delta q = \sum X_i\, (1 + Z_i)\, \delta n_1^i \,\,\, - &
 \sum X_i\, (1 + Z_i)\, Z_i\, \delta n_2^i \,- \, \delta s_1\, \sum X_i\, (1 + Z_i) \,+
 \nonumber \\
 & \delta s_2\, \sum X_i\, (1 + Z_i)\, Z_i\, , \label{eqn:diff} 
\end{eqnarray}
where
\begin{eqnarray}
X_i = Z_i\, \cos 4\psi_i - q \,(1 + Z_i)\, \cos^2 4\psi_i -
u\, (1 + Z_i)\, \sin 4\psi_i\, \cos 4\psi_i \nonumber
\end{eqnarray}
and $n_*$ is the counts corresponding to the Stokes parameter $I$, the
intensity due to the object alone.

The gain ratio of the beams
$\alpha$ is usually  close to unity  and $q$ $\ll$ 1, 
making the counts $n_1^i$'s and
$n_2^i$'s, and the statistical errors in their determination
similar. Since photons obey Poisson statistics, we have $\sigma_n
= \sqrt{n}$, in general. Assuming $\sigma_{n_1^i} = \sigma_{n_2^i}
= \sigma_{n}$  and $\sigma_{s_1}
= \sigma_{s_2} = \sigma_{s}$ ,
using the law of propagation of errors, from Equation~\ref{eqn:diff} we get
\begin{eqnarray}
(\sigma_q)^2 = \frac{4}{M\, n_*^2} \left\{(\sigma_n)^2 + 
2.5\, M\, q^2\, (\sigma_s)^2\right\}\, .  \nonumber
\end{eqnarray}
If $\bar{n}$ and $\bar{s}$, respectively, are the averages of the observed
object $plus$ background sky and
background sky counts at various positions of the rotating half-wave plate
in one second and if the object is observed for
$t_o$ and background for $t_b$ seconds at each position of the half-wave
plate, then
\begin{eqnarray}
 & n_* = 2\,(n - s) = 2\,t_o\,(\bar{n} - \bar{s})\, ,& \nonumber \\
& (\sigma_n)^2 = t_o\,\bar{n} \quad \mbox{and} \quad 
(\sigma_s)^2 = (t_o/M\,t_b)^2\,M\,t_b\,\bar{s}  = 
t_o^2 \bar{s}/M\,t_b\, . & \nonumber
\end{eqnarray}
Making use of these values we get the probable error in $q$ as
\begin{eqnarray}
\epsilon_q = \frac{0.6745}{\sqrt{M}\,(\bar{n} - \bar{s})}
\left(\frac{\bar{n}}{t_o} + 2.5\,q^2\,\frac{\bar{s}}{t_b}\right)^{\frac{1}{2}}.
\nonumber
\end{eqnarray}
There will be a similar relation for the other 
normalized Stokes parameter $u$.
It is clear from the above that the errors in $q$ and $u$ due to statistical
fluctuations in the accumulated counts depend on the values
of $q$ and $u$ themselves.
When $\epsilon_q = \epsilon_u$, $\epsilon_p = \epsilon_q$.
According to the above relation for the error
due to photon noise $\epsilon_q = \epsilon_u$, when
$q$ = $u$, i.e.,  when
$q$ = $p/\sqrt{2}$. This implies that for a given
linear polarization $p$, the observations would yield a minimum error in
$p$  if $q$ = $u$, all other parameters being the same.
The highest possible error in P due to photon fluctuations,
which occurs when either $q$ = $p$
and $u$ = 0, or $u$ = $p$ and $q$ = 0,
is obtained by replacing $q$ by $p$ in the above equation as
\begin{eqnarray}
\epsilon_P(\%) = \frac{67.45}{\sqrt{M}\,(\bar{n} - \bar{s})}
\left(\frac{\bar{n}}{t_o} + 2.5\,p^2\,\frac{\bar{s}}{t_b}\right)^{\frac{1}{2}}\, .
\nonumber
\end{eqnarray}

\begin{figure}
\centerline {\includegraphics[width=9cm]{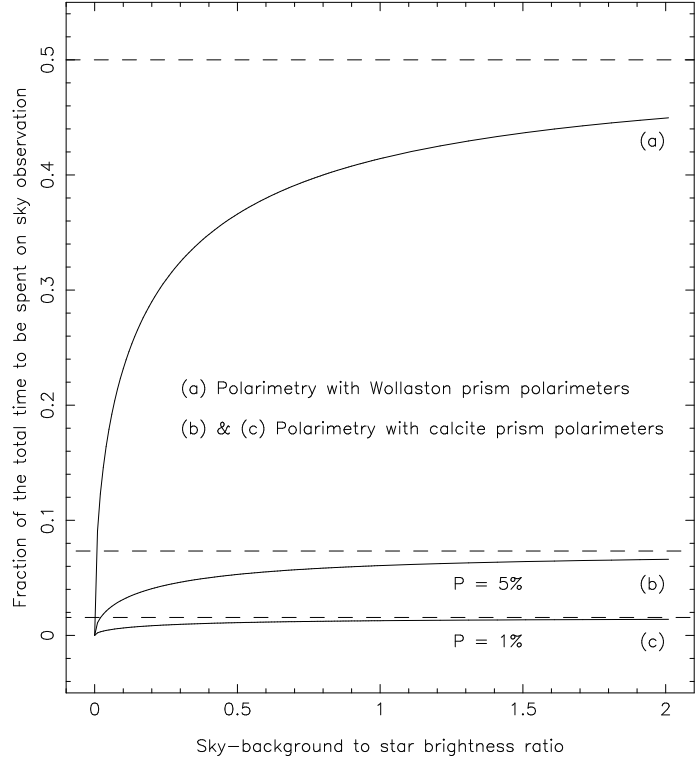}}
\caption{Plot of the optimum fraction of the time to be spent on background
sky observation
with  the beam-displacement- and Wollaston prism-based polarimeters against the
sky-background to star brightness ratio.
The dashed lines show the corresponding asymptotic values.}
\label{f:opt}
\end{figure}

For an efficient use of the telescope time, we should distribute the total
time available between the object and background sky observations such that the 
signal-to-noise ratio is a maximum, or equivalently, the error in P is 
a minimum. Differentiating the expression for ($\epsilon_P$)$^2$ with
respect to $t_o$, the time spent on object integration, and equating it to zero,
we get the condition
for $\epsilon_P$ to be a minimum as
\begin{equation}
\frac{t_b}{t_o} = p\,\sqrt{\frac{2.5\,\bar{s}}{\bar{n}}}\, .
\end{equation} 

In polarimeters where the beam separation is large,
the moon lit background sky will be modulated by the rotating half-wave plate
because of its polarization. The intensities of the
ordinary- and extraordinary-components
at each position of the half-wave plate
should be measured separately
and removed from the object $plus$ background data. When we deal with
such a situation,
$s_1$ and $s_2$ appearing in the above equations should be replaced by
the corresponding $s^i_1$'s and $s_2^i$'s, which
have uncorrelated statistical errors.
The equation for ($\epsilon_P$)$^2$ then would be modified to
\begin{eqnarray}
\epsilon_P(\%) = \frac{67.45}{\sqrt{M}\,(\bar{n} - \bar{s})}
\left(\frac{\bar{n}}{t_o} + \frac{\bar{s}}{t_b}\right)^{\frac{1}{2}},
\nonumber
\end{eqnarray}
and the condition
for $\epsilon_P$ to be a minimum as
\begin{eqnarray} 
\frac{t_b}{t_o} = \sqrt{\frac{\bar{s}}{\bar{n}}}\, .
\nonumber
\end{eqnarray}
Fig.~\ref{f:opt} 
shows the fraction of the time to be spent on background sky
observations
as a function of the ratio of the background brightness to the object
brightness with two different types of polarimeters, one with
well separated ordinary- and extraordinary-beams as in the case
of Wollaston or Foster prism-based polarimeters, and the other
with overlapping ordinary- and extraordinary-beams as in the
case of beam-displacement prism-based polarimeters. In the case
of overlapping beam polarimeters the fraction of time to be spent
on background observation increases with the polarization of the
object and at low polarization levels, which is usually encountered
in stellar polarization measurements, it is negligibly small,
indicating that most of the available time can be spent on
observing the object, and thereby, partially compensating for the loss
in efficiency in not utilizing 50 per cent of the
light collected by the telescope.

\section{Observational validation}
The two main parameters of a polarimeter that determine its suitability
for observations are the polarization it produces for an unpolarized beam
and its ability to measure correctly the degree of
polarization of a polarized beam without causing any depolarization;
during February--April~2015 we observed unpolarized stars with and without
the Glan-Taylor prism in the
telescope beam to determine these two.
Observations during April--May~2014 were made
with different settings of the
chopper speed, number of positions of the
halfwave plate over a full rotation, integration times
and diaphragm sizes, in order to look for any dependency on their values.
Since we could not find any obvious differences in the results of those
observations,
all the observations made during February--April~2015, which are discussed
here, were done with a
chopper frequency of 50~Hz, 25 positions of the halfwave plate
over a full rotation,
integration time of 1-s, and diaphragms of 20~arc sec diameter.
In addition to the $UBVRI$ bands, another broad spectral band which included
both the $R$ and $I$ bands was also used for the observations. We refer to 
this band, which has a mean wavelength of 712~nm, in this paper as $R'$.

\begin{figure}
\centerline {\includegraphics[width=9.0cm]{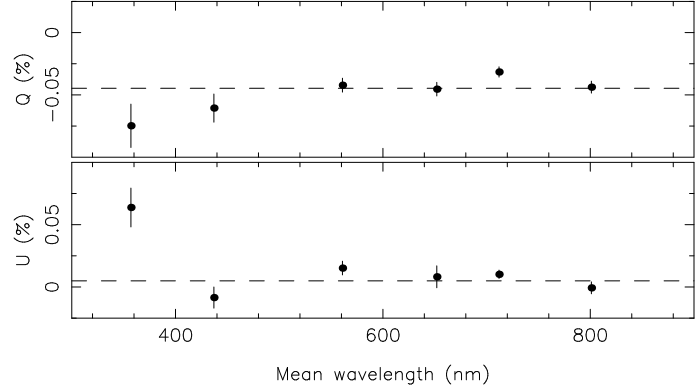}}
\caption {Plots of instrumental Q (\%) and U (\%) against the mean wavelength
of the corresponding spectral band. The dashed lines show the averages of the
corresponding quantities in the $BVRR'I$ bands.
\label{f:instrqu}}
\end{figure}

\begin{figure}
\centerline {\includegraphics[width=9.0cm]{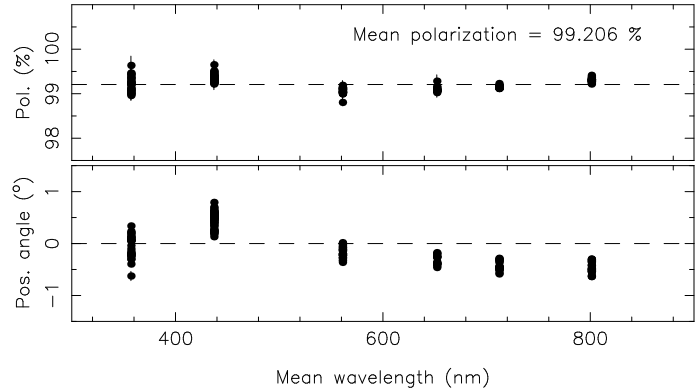}}
\caption {Plots of polarization efficiency and position angle
against the mean wavelength
of the corresponding spectral band. The dashed lines show the 
corresponding mean values.
\label{f:poleff}}
\end{figure}

\subsection{Instrumental polarization}
We observed the unpolarized stars, HD~42807, HD~65583, HD~90508, HD~98281, 
HD~103095, HD~100623, HD~125184 and HD~144287, on several occasions.
The average values of the observed Q (\%) and U (\%)
in the $UBVRR'I$ bands are plotted against the mean wavelengths of the
spectral bands in Fig.~\ref{f:instrqu}.
It is clear that the polarization produced by the
telescope-polarimeter combination, which is usually referred to as 
the instrumental polarization, is small. The instrumental polarization,
apparently, increases slightly towards the ultraviolet. It is nearly
constant in the $V-I$ spectral region.

\begin{figure}
\centerline{\includegraphics[width=7.7cm]{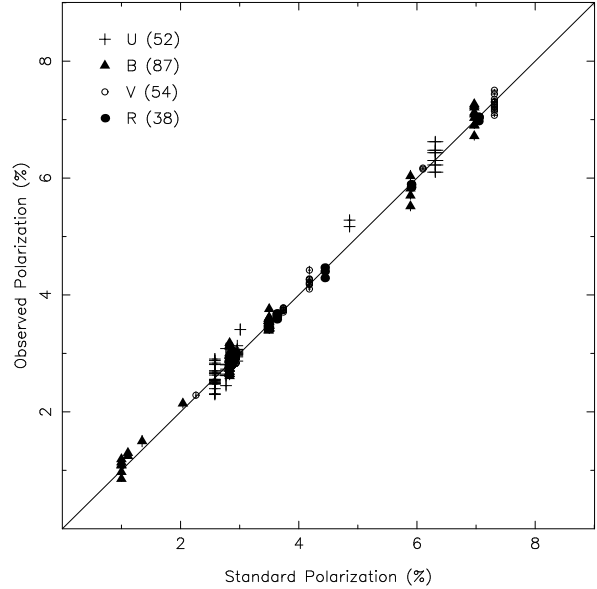}}
\caption {Plot of observed polarization against that available
in the literature.
The observed polarization has been corrected for the wavelength-dependent
polarization efficiency and the instrumental polarization.
The straight line indicates an efficiency of 100 per cent for the polarimeter.
The number inside the brackets indicates the number of observations
obtained in the corresponding spectral band.
\label{f:stan_pol}}
\end{figure}

\subsection{Polarization efficiency}
A total of 161 observations of several unpolarized stars were made with the
Glan-Taylor prism in the light path of the telescope beam to determine the
polarization efficiency of the instrument, which is the numerical
value obtained by the instrument for an input beam that is 100\% polarized.
In the top panel of Fig.~\ref{f:poleff} we have plotted the individual
values of the polarization efficiencies obtained by us in $UBVRR'I$ spectral
bands against the corresponding mean wavelength. The polarization efficiency
has a slight wavelength dependence, with lower values in the $V-R$ spectral
region. The mean polarization efficiency is 99.206\% and
its total amplitude of variation in the $U-I$ spectral region is 0.271\%.
The position angle of polarization observed, which is plotted in the
bottom panel of Fig.~\ref{f:poleff}, also shows a slight wavelength dependence. 
The wavelength dependence observed is almost an inverted and
scaled-down version of
the variation of the position angle of
the effective optical axis theoretically computed for a super-achromatic
halfwave plate \citep{serk2}, indicating that the fixed super-achromatic
halfwave plate in the beam does not fully compensate for the variation in
the position angle of the effective optical axis of the rotating plate
because of the slight, but unavoidable errors in their fabrication.
The total amplitude of variation in the position angle is only 0.92\degr \, and
the wavelength-dependent offset in the position angle
can be incorporated in the data reduction procedure easily.

\begin{figure}
\centerline{\includegraphics[width=7.7cm]{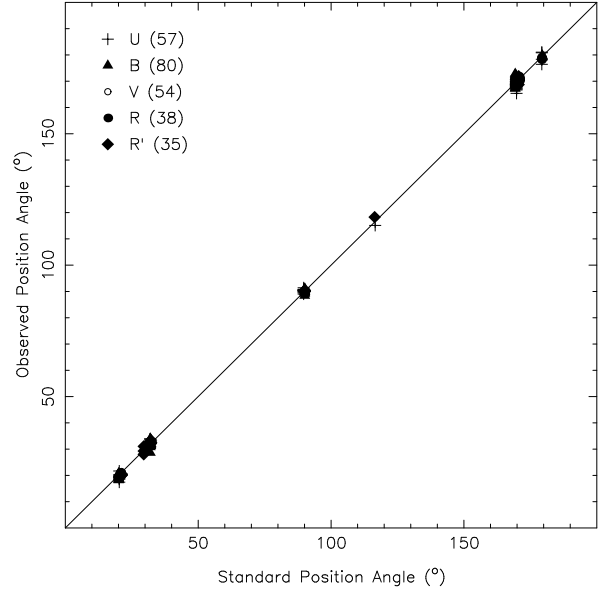}}
\caption {Plot of observed position angle against that available
in the literature. The observed position angle has been corrected for the
wavelength-dependent offset. The straight line has a slope of unity.
The number inside the brackets indicates the number of observations
obtained in the corresponding spectral band. \label{f:stan_ang}}
\end{figure}

\subsection {Observations of polarized stars}
During the observing runs we
observed the polarized stars, HD~21291, HD~23512, HD~43384, HD~147084, 
HD~154445, HD~160529, HD~183143, HD~58439,
HD~94473, HD~127769 and HD~142863 in $UBVRR'I$ bands. Most of the above
objects were observed several times during the observing runs.
The first 7 objects are considered to be standard polarized stars, and
are normally used to determine the offset in the measured position angles
from the standard equatorial coordinate system.
\citet{hsu} have reported polarizations and position angles for
these objects in $UBVR$ bands. For the other 4 stars, 
\citet{math} have given polarization measurements in the $B$ spectral band.
The above 4 stars were included in the 
present observations so as to have an extended range in the
polarization and brightness for the observed polarized stars.
In Fig.~\ref{f:stan_pol} 
we have plotted the polarization determined by us in $UBVR$
bands against the corresponding value available in the literature. The
observed polarization has been corrected for the wavelength-dependent
polarization efficiency of the instrument.
It is clear from Fig.~\ref{f:stan_pol} that there is an
excellent agreement between the measured polarization
and the corresponding value available in the literature.
A linear least square fit to the data plotted in the figure
gives a value of 1.004$\pm$0.001 for the slope.

We have plotted in Fig.~\ref{f:stan_ang}
the position angles observed by us in $UBVRR'$ against the
corresponding values available in the literature.
The position angles in $R'$ band were obtained by an interpolation of the
data given in \citet{hsu}.
It is clear that the agreement between the measured values and those available
in the literature is very good.
The observed position angles were corrected for the wavelength-dependent
position angle of the effective optical axis of the
rotating halfwave plate. The offset in the position angles was 
determined by a least square fit to the combined data plotted
in Fig.~\ref{f:stan_ang}. The offset obtained using
such a procedure would be better than the
value determined using a single standard polarized star,
since, the position angles of some of the standards are found to vary by
more than 1\degr \, \citep{hsu}.

\begin{figure}
\centerline{\includegraphics[width=7.7cm]{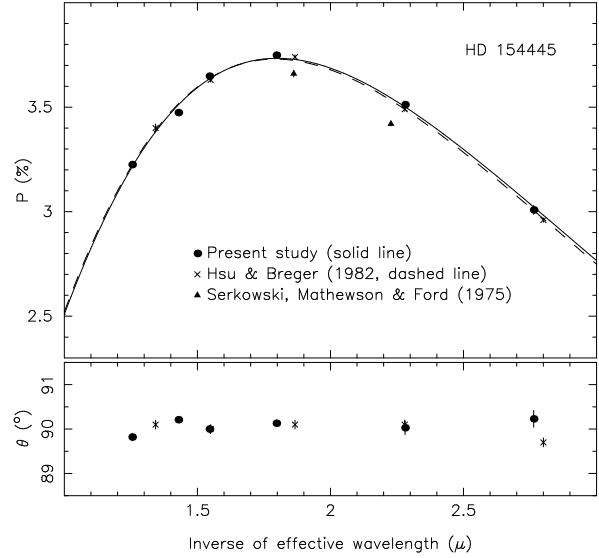}}
\caption {Plot of polarization in $UBVRR'I$ bands observed against the
inverse of the corresponding effective wavelength. The solid 
($P_{max} = 3.73\pm0.01$, $\lambda_{max} = 556\pm$2 nm) and dashed
($P_{max} = 3.73\pm0.01$, $\lambda_{max} = 558\pm$2 nm) lines
show the computed interstellar polarization curves using the data obtained
by us and \citet{hsu}, respectively.
\label{f:hd154445}}
\end{figure}

In Fig.~\ref{f:hd154445} we have plotted the averages of the polarization
observed in $UBVRR'I$ bands against the corresponding inverse
of the effective wavelength,
which were computed using the responses of the filter-detector combinations,
the $UBVRI$ magnitudes in the Johnson's system and the average
extinction coefficients in those bands observed at Kavalur, for one of the
polarization standards, namely, HD~154445. The airmass of observation
was taken as 1.5. 
The figure also shows the observations of \citet{hsu} and \citet{serk}; the
effective wavelengths of these observations
were computed using the empirical relations given by the respective
authors. The airmass of observation was taken as 1.0, while computing
the effective wavelengths of the observations of \citet{hsu}.
The position angle of polarization of the object
is plotted in the bottom panel of the figure.

We have also plotted in Fig.~\ref{f:hd154445}
the interstellar polarization curves, computed using the
empirical relation given by \citet{serk}.
The $P_{max}$, the maximum value of polarization, and $\lambda_{max}$,
the wavelength at which the maximum occurs, of the present data were
derived using a least square fit of the empirical relation. \citet{hsu}
have given $P_{max}$ and $\lambda_{max}$ for their data.
It is clear from the figure that the present observations
are in excellent agreement with those of \citet{hsu}.

\section{Conclusions}
A new polarimeter for simultaneous observations in
three spectral bands, $U$, $B$ and $V$, or $R$, or $I$ was designed and built 
in the Indian Institute of Astrophysics. A cross-mounted Calcite beam-
displacement prism acts as the analyzer, and the modulation of the
intensities of the
emergent beams is achieved by a rotating superachromatic halfwave plate.
Combinations of dichroic and Schott glass filters are used to isolate
the spectral bands. In each spectral band
the emergent beams from the analyzer are quasi-simultaneously detected by
the same photomultiplier tube using a fast rotating chopper.
The operation of the polarimeter is done using a Linux machine. All the
functions of the polarimeter are controlled by the electronics built
around PIC microcontrollers.

Since the background sky is not polarized when a
beam displacement prism is used as the analyzer, 
time has to be spent only to
determine the brightness of the background sky. We show from
an analysis of the propagation of errors that the time to be
spent on background observation is only a small fraction of the object
integration for an optimum 
error in the observed polarization, which arise from
the statistical fluctuations in the photon counts. This advantage partially
offsets the loss in the efficiency of beam displacement prism-based
polarimeters in not utilizing 50\% of the light collected by the
telescope.

The polarimeter was mounted onto the 1-m Carl Zeiss telescope at Vainu Bappu
Observatory, and several unpolarized and polarized stars
were observed during February--April~2015 to test its suitability for efficient
astronomical observations. An analysis of the data collected
showed the performance of the polarimeter to be quite satisfactory.
The instrumental polarization, which includes the telescope contribution also,
is found to be small. Observations with the Glan-Taylor prism in the light
path show that both
the polarization efficiency and the position angle of polarization are
slightly wavelength-dependent; however, the total amplitudes of variation
in the $U-I$ spectral region are only 0.271\% and 0.92\degr \,.
The mean polarization efficiency is found to be 99.206\%.
Both the polarization
and position angles of standard polarized stars obtained using the polarimeter
are in excellent agreement with those available in the literature.

\section*{Acknowledgments}
We gratefully acknowledge the keen interests shown by 
Professors H.~C.~Bhatt and P.~Sreekumar, and
Shri~A.~V.~Ananth in putting the instrument in 
operation at the telescope.
We thank Professors G.~V.~Coyne, F.~Scaltriti, and A.~M.~Magalhaes
for their prompt responses to our queries, which helped us in
finalizing the optical layout, and
Professor~F.~Scaltriti for
sending copies of the drawings of the polarimeter at Torino
Astronomical Astronomical Observatory.
Several of our colleagues helped us at various stages;
Mr~P.~K.~Mahesh, Mr~P.~M.~M.~Kemkar, Mr~P.~U.~Kamath,
Dr~D.~Suresh and Dr~G.~Rajalakshmi helped us in the preparation of the
Auto CAD drawings of the mechanical parts of the polarimeter;
Professor T.~P.~Prabhu helped us in acquiring the dichroic mirrors and
glass filters;
Mr~N.~Sivaraj helped us in checking the photomultiplier
tubes and the pulse-amplifier-discriminators; the staff at
Vainu Bappu Observatory helped us in carrying out the
observations using the instrument; we thank all of them.
We thank Professor T.~P.~Prabhu also for critically going through the
manuscript and making valuable suggestions for improving the quality of
the paper.

\label{lastpage}

\end{document}